\documentclass[floats,twocolumn]{revtex4}

\usepackage{graphicx}
\usepackage{amsfonts}

\newcommand{\DeclareMathOperator}[2]{\def#1{\mathop{\mathrm{#2}}\nolimits}}

\DeclareMathOperator{\Tr}{Tr}
\DeclareMathOperator{\tr}{tr}

\DeclareMathOperator{\sgn}{sgn}

\def\be{\begin{equation}}
\def\ee{\end{equation}}
\def\ba#1{\begin{array}{#1}}
\def\ea{\end{array}}
\def\bn{\begin{enumerate}}
\def\en{\end{enumerate}}

\def\beq{\begin{equation}}
\def\eeq{\end{equation}}

\begin{document}

\title{Entanglement spectrum of topological insulators and superconductors}
\author{Lukasz Fidkowski}
\affiliation{California Institute of Technology, Pasadena, CA 91125, U.S.A.}

\begin{abstract}
We study two a priori unrelated constructions: the spectrum of edge modes in a band topological insulator or superconductor with a physical edge, and the ground state entanglement spectrum in an extended system where an edge is simulated by an entanglement bipartition.  We prove an exact relation between the ground state entanglement spectrum of such a system and the spectrum edge modes of the corresponding spectrally flattened Hamiltonian.  In particular, we show that degeneracies of the entanglement spectrum correspond to gapless edge modes.
\end{abstract}

\maketitle

\section{Introduction \label{intro}}

Topological phases of matter, occurring for example in the quantum Hall effect, cannot be distinguished using a local order parameter.  Although they can sometimes be characterized using, for example, topological ground state degeneracy or the existence of gapless edge modes, a full understanding is still lacking.  One worthwhile direction is to use the information-theoretic concept of entanglement to characterize topological phases, as was shown first in \cite{kp,lw}, who related the universal sub-leading term in the bipartite entanglement entropy (the ``topological entanglement entropy") to the total quantum dimension in a gapped anyonic system.  Further progress was made by Haldane and Li \cite{haldane} who related the full spectrum of the reduced density matrix (the ``entanglement spectrum") to the conformal field theory (CFT) edge mode spectrum in fractional quantum Hall states.  Here, our goal will be to study topological phases occurring in free fermion systems, i.e. band topological insulators and superconductors.  In such systems the topological entanglement entropy can vanish, so one needs a finer method to discriminate between the phases.  We will adhere to the philosophy of \cite{haldane} and focus on the entanglement spectrum, which we will show does contain more information about the various free fermion topological phases.  Specifically, we will show that the entanglement spectrum of any band insulator or superconductor can be exactly reconstructed from the edge mode spectrum of the corresponding spectrally flattened Hamiltonian.

Given a division of a quantum system into two subsystems $A$ and $B$, such that the total Hilbert space is the tensor product of the subsystem Hilbert spaces, one defines the reduced density matrix $\rho_A$ on $A$ by tracing out the degrees of freedom of $B$ from the pure ground state density matrix $| \psi \rangle \langle \psi |$ (for fermion Hilbert spaces one actually has to introduce graded tensor products, but we can safely ignore this subtlety).  The von Neumann entanglement entropy between $A$ and $B$ relative to the ground state is then defined by $S = - \tr \rho_A \log \rho_A$.  This entropy is a measure of the complexity of the ground state, as seen, for example, in the logarithmic scaling of entanglement entropy at criticality \cite{klich}, as well as in the study of one dimensional gapped systems, where finite entropy matrix product states (MPS) are introduced to approximate ground states \cite{vers, hastings}.

As motivated by \cite{haldane}, we take the more general approach of studying the entire spectrum of eigenvalues of the reduced density matrix (i.e. the eigenvalues of the Schmidt decomposition of the ground state), in the hope that the spectrum contains more information than just the one number $S$ that can be constructed from it.  We restrict our attention to free fermion systems, which, though non-interacting, include effective Hamiltonians for topological insulators, such as HgTe in $2$ dimensions \cite{KaneMele1, HgTe0} and BiSb in $3$ dimensions \cite{BiSb0, BiSb1}, as well as superconductors such as SrRu, and systems with broken $T$ symmetry such as the integer quantum Hall effect.  

There has been a great deal of work done on entanglement in free fermion systems.  First of all, entanglement in a Fermi gas is well understood \cite{klich, levitov}.  The generalization to arbitrary free fermion systems has been studied as well \cite{peschel1, peschel2, uli}, and the problem of computing the entanglement spectrum has been reduced to diagonalizing a matrix of Green's functions.  Here we derive this formula in the formalism of free Majorana fermions and Gaussian states \cite{kitaev, bravyi}.  The advantage of this approach is that it unifies the treatment of both band topological insulators and superconductors in that the effective Hamiltonian, which can include both hopping and pairing terms, becomes a general quadratic form in the Majorana fermions.  We use this formula to relate the entanglement spectrum to gapless edge modes of a sample with boundary.  Before we go into more detail, we make some comments on related work.

In \cite{nba}, a numerical approach was pursued in the study of the entanglement spectrum of a topological superconductor.  Specifically, it was shown numerically that one can meaningfully distinguish between the weak and strong pairing phases of a $2$ dimensional $p+ip$ superconductor by looking at the degeneracies of the entanglement spectrum.  In this paper we analytically prove a relation which generalizes this result.  Furthermore, in \cite{turner-2009}, which appeared shortly after this paper, a general relation similar to ours was derived.  Working in the setting of topological insulators, \cite{turner-2009} show that gapless edge modes imply degeneracies in the entanglement spectrum.  However, they also show that the converse is not true by exhibiting a specific model of a $3$ dimensional topological insulator in which a Zeeman field gaps out the edge modes but does not effect the degeneracies of the entanglement spectrum, which are protected by inversion symmetry.

Let us now describe our derivation in more detail.  We prove a relation between two a priori unrelated constructions.  The first is the computation of the entanglement spectrum of the ground state of a band topological insulator or superconductor with respect to a partition into a large but finite region $A$ (such as a disc in $2$ dimensions, for example), and its complement $B$.  For the second, we need the `spectral flattening trick' \cite{kitaev}, which we use to deform the original Hamiltonian $H$ to a new one $H'$ that retains the same ground state, but has a flat spectrum; this is always possible for a gapped Hamiltonian.  The deformation keeps the gap open and can be done adiabatically, showing that the two Hamiltonians are in the same topological phase.  We now simulate a physical edge at the boundary of $A$ by defining $H'_A$ to be the restriction of the spectrally flattened Hamiltonian to region $A$ - that is, we retain only the couplings within $A$, and discard the degrees of freedom in the now decoupled region $B$.  We prove that the spectrum of this restricted, spectrally flattened, Hamiltonian $H'_A$ and the entanglement spectrum of the ground state can be reconstructed from each other.

Of particular interest are the low energy modes of $H'_A$, which, because the bulk is gapped, are edge modes localized at the boundary of $A$.  One consequence of our formula is that zero modes of $H'_A$ correspond to degeneracies in the entanglement spectrum.  An important point, observed in \cite{turner-2009}, is that the edge modes of $H'_A$ might not be the same as those obtained from the restriction $H_A$ of the original Hamiltonian to $A$, without first applying a spectral flattening transformation.  Indeed, as was shown in \cite{turner-2009}, it is sometimes the case that the edge modes of $H_A$ are gapped, even though there are degeneracies in the entanglement spectrum (protected by inversion symmetry, for example).  The crux of the issue, thus, is that the edge mode spectrum can change during the spectral flattening transformation.  However, if there is a symmetry, such as time reversal, that protects the gapless nature of the edge modes, then all the Hamiltonians along the spectral flattening deformation respect this symmetry, and possess gapless edge modes.  In this case, therefore, we can conclude that there are at least as many corresponding degeneracies in the entanglement spectrum, in a sense made precise below.  As \cite{turner-2009} propose, the entanglement spectrum may in fact be a more robust characterization of a topological phase than the edge mode spectrum (see also \cite{pollmann-2009}).

\begin{figure}[htp]
\includegraphics[width=6cm]{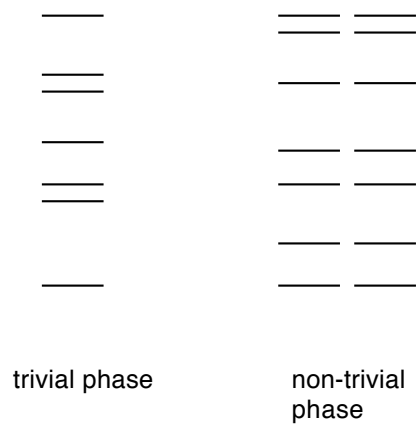}
\caption{Schematic representation of the entanglement spectrum of the Majorana chain.  In the nontrivial topological phase, there is a two-fold degeneracy in the spectrum. \label{es1}}
\end{figure}

Before we move to the general analysis, we illustrate our point with a simple example.  Consider a one dimensional superconductor, i.e. a fermionic Hamiltonian with both pairing and hopping terms.  The Hamiltonian is: \begin{equation} H=\frac{1}{2} \, \sum_{l} \left( u \, a_l a_l^\dagger + v \, a_l a_{l+1} + v \, a_l^\dagger a_{l+1} \right) + \text{h.c.}. \end{equation}  This is the so-called Majorana chain Hamiltonian \cite{kitaev-2000}, and it exhibits two phases.  For $|u|>|v|$, the chemical potential dominates, and in the limit of large $u/v$ simply forces each site into occupation number $0$ or $1$, depending on the sign of $u$.  At $|u|=|v|$ there is a critical point describing a transition into a different phase ($|u| < |v|$) (in fact, a Jordan-Wigner transform maps this to the transverse field Ising model transition).  This new phase is topological, in the sense that it is characterized by having unpaired Majorana modes at the boundary \cite{kitaev-2000}.  That is, for chain whose length $L$ is much larger than the correlation length $\xi$, there are two ground states, degenerate up to a splitting $\exp(-L/\xi)$.  They form a two state system obtained from pairing up the boundary Majorana modes.  However, we will find below that the entanglement spectrum also contains a signature of the topological phase.  Namely, we will see that in the topological phase, the multiplicity of all eigenvalues in the Schmidt decomposition is doubled (see fig. \ref{es1}).

\section{General Analysis \label{gen}}

We consider a free fermion system (i.e. band insulator or superconductor) in an arbitrary number of dimensions $d$.  We picture a tight binding model with short range interactions, which could be either hopping or pairing.  To conveniently work with both, we write each orbital in terms of two Majorana fermions, so that $N$ physical fermion modes are described with $2N$ operators $c_j$, $j=1, \ldots, 2N$.  The $c_j$ are Hermitian and satisfy the Majorana commutation relations $\{ c_j, c_k \} = 2 \delta_{jk}$.  We can write $N$ physical fermion creation and annihilation operators as \begin{eqnarray} a_n &=& \frac{1}{2} \left( c_{2n} + i c_{2n-1} \right) \\ a_n^\dagger &=& \frac{1}{2} \left( c_{2n} - i c_{2n-1} \right). \end{eqnarray} The effective Hamiltonian can now be written in terms of the $c_j$: \begin{equation} \label{gh} {H = \frac{i}{4} \sum_{j,k=1}^{2N} H_{jk} \, c_j c_k}, \end{equation} where we interpret $j$ as a collective site and band index.  The matrix $H_{jk}$ is real skew-symmetric, and we assume the Hamiltonian (\ref{gh}) is gapped.

We now choose a partition of the system into a subsystem $A$ and its complement $B$.  For instance, in two dimensions we could define $A$ and $B$ by partitioning the lattice into two complementary half-planes.  However, for convenience we shall assume $A$ to contain finitely many lattice sites.  Indeed, let $A$ contain $m$ orbitals, described by the Majoranas $c_1, \ldots, c_{2m}$, and let $B$ be described by the remaining Majoranas $c_{2m+1}, \ldots, c_{2N}$.

\subsection{Gaussian States}
Before proceeding, it will be useful to introduce the concept of a Gaussian state \cite{bravyi}.  Intuitively, a Gaussian state is simply the formalization of what it means for a (possibly mixed) state to be the ground state of a free fermion Hamiltonian.  Since free fermion Hamiltonians can be diagonalized, one can think of a Gaussian state of as a tensor product of independent $2$-state system density matrices in some orthogonal basis.  More formally, consider a density matrix $\rho$, written as a polynomial of the $2N$ Majoranas $c_j$ in such a way that each $c_j$ occurs to the power $0$ or $1$ in each term.  The state defined by $\rho$ is said to be Gaussian if, upon replacing the $c_j$ with anti-commuting Grassman variables $\theta_j$, one obtains an expression ${\bar \rho}$ that can be put in the form \begin{equation} \label{ef} {\bar \rho} = \frac{1}{2^{N}} \exp \left( \frac{i}{2} \theta^T M \theta \right) \end{equation} for some real antisymmetric $2N$ by $2N$ matrix $M$.  The matrix $M$ simply encodes the $2$ point correlators of the $c_j$ in the state $\rho$: \begin{equation} \label{cor} M_{jk} = \Tr (\rho \, i c_j c_k) \end{equation} for $j \neq k$, with $M_{jj}=0$.  All higher correlators are determined by Wick's theorem.


The states with which we will be dealing are all Gaussian.  First of all, the ground state $|\psi \rangle \langle \psi|$ of a gapped Hamiltonian is Gaussian (to see this, simply bring $H_{jk}$ to canonical block diagonal form).  Furthermore, given any Gaussian state $\rho$ of the full system, the reduced density matrix $\rho_A$ constructed from it by tracing out the degrees of freedom in $B$ is also Gaussian.  Indeed, because the correlators of $\rho_A$ are the same as those of $\rho$, and a state is determined uniquely by the set of all its correlators, we see that $\rho_A$ is a Gaussian state whose matrix $M$ (\ref{ef}) is simply the restriction of that of $\rho$.

What is the $M$ matrix (\ref{ef}) for the ground state of the Hamiltonian (\ref{gh})?  We can determine it from the $2$ point functions of the ground state, obtained from the formula \cite{kitaev} \begin{equation} \label{ke}  \langle \psi | c_j c_k | \psi \rangle =  -\sgn \left( i\, H_{jk} \right). \end{equation}  The $\sgn$ function is defined as follows: for a diagonal Hermitian matrix $D$, it replaces all positive eigenvalues with $+1$ and all negative eigenvalues with $-1$.  A general Hermitian matrix, such as $i \, H$ in (\ref{ke}), can be diagonalized with a unitary transformation $U$: $i \, H = U \, D \, U^{-1}$, and we define $\sgn \left( i\, H_{jk} \right) = U \,  \sgn \left( D \right) \, U^{-1}$.  Thus, after comparison to eq. (\ref{cor}), we see that \begin{equation} \label{msh} M_{jk} = -i \sgn (i \,H_{jk}). \end{equation}

\subsection{Exact correspondence}

Having introduced the necessary formalism, we now prove an exact relation between the entanglement spectrum of a general gapped Hamiltonian, and the edge mode spectrum of the corresponding spectrally flattened one.  Let us first explain the spectral flattening transformation \cite{kitaev}.  Given any gapped Hamiltonian (\ref{gh}) we can construct a one parameter family of gapped Hamiltonians that interpolate between $H$ and \begin{equation} H' = \frac{i}{4} \sum_{j,k=1}^{2N} H'_{jk} \, c_j c_k, \end{equation} where the eigenvalues of $i\, H'_{jk}$ are all $\pm 1$.  The matrices that interpolate between $H_{jk}$ and $H'_{jk}$ share a common eigenbasis; they are all gapped, and leave the ground state invariant.  Also, as a consequence of the gap being open, all of the Hamiltonians in the family are quasi-local, with quadratic coupling terms exponentially suppressed by the distance \cite{kitaev}.

Because the spectral flattening transformation leaves the ground state invariant, the entanglement spectrum, which depends only on the ground state, is the same for $H$ and $H'$.  Thus, to prove our relation we will from now on just assume $H$ is spectrally flat, and omit the extra superscript in $H'$.  With this assumption, equation (\ref{msh}) simplifies to \begin{equation} \label{mh} M_{jk} = H_{jk}. \end{equation}

Equation (\ref{mh}) is the key to proving the correspondence.  Roughly speaking, we will relate its right hand side to the edge mode spectrum, and its left hand side to the entanglement spectrum.  More formally, let ${\tilde H}_{jk}$, and ${\tilde M}_{jk}$ denote the restrictions of $H_{jk}$ and $M_{jk}$ to $A$ respectively; thus ${\tilde H}_{jk}$ and ${\tilde M}_{jk}$ are $2m$ by $2m$ anti-symmetric matrices, and they are equal by virtue of (\ref{mh}).  The matrix ${\tilde H}_{jk}$ now defines a physical Hamiltonian on region $A$: \begin{equation} \label{ha} H_A = \frac{i}{4} \sum_{j,k \in A} {{\tilde H}_{jk}} \, c_j c_k. \end{equation}  $H_A$ is gapped in the bulk, with spectral gap normalized to $1$ in our units, but it also has boundary modes, which could potentially be gapless.  The eigenvalues of $i \, {\tilde H}_{jk}$, which come in pairs $\pm \lambda_r$, $|\lambda_r| \leq 1$, $r = 1, \ldots, m$, reflect these boundary modes, in that the corresponding eigenstates are localized near the boundary whenever $\lambda_r$ differs substantially from $\pm 1$.



From our discussion of Gaussian states, on the other hand, we see that ${\tilde M}_{jk}$ is just the $M$ matrix (see eq. \ref{ef}) of $\rho_A$, the reduced density matrix obtained from the ground state $| \psi \rangle \langle \psi |$ by tracing out the degrees of freedom in $B$.  This means that if one defines ${\bar \rho}_A$, as in (\ref{ef}), to be $\rho_A$ with the Majoranas $c_j$, $j=1, \ldots, 2m$, replaced with anti-commuting Grassman variables $\theta_j$, then \begin{equation} \label{era} {\bar \rho}_A = \frac{1}{2^m} \exp \left( \frac{i}{2} \theta^T \tilde{M} \theta \right) = \frac{1}{2^m} \exp \left( \frac{i}{2} \theta^T \tilde{H} \theta \right). \end{equation} Since the expression $\theta^T \tilde{H} \theta$ is $SO(2m)$ invariant, we can rotate to a more convenient basis.  In particular, let ${\tilde H}^c$ be the canonical block diagonal form of ${\tilde H}$, consisting of $m$ blocks of the form \begin{equation} \label{can} \left( \begin{array}{cc} 0 & \lambda_r \\ -\lambda_r & 0 \end{array}\right). \end{equation} Let $M$ be the $SO(2m)$ matrix that block diagonalizes ${\tilde H}$: ${\tilde H}^c = M \, \tilde{H} \, M^{-1}$, and define $\theta'_j = \sum_k M_{jk} \, \theta_k$, and $c'_j = \sum_k M_{jk} \, c_k$.

In terms of the rotated variables, (\ref{era}) turns into \begin{eqnarray} \label{fe} \nonumber {\bar \rho}_A &=& \frac{1}{2^m} \exp \left( \frac{i}{2} {\theta'}^T \, \tilde{H}^c \, \theta' \right) \\ \nonumber &=& \frac{1}{2^m} \prod_{r=1}^m \exp \left( i\, \lambda_r \, \theta'_{2r-1} \theta'_{2r} \right) \\ &=& \prod_{r=1}^m \left(\frac{1}{2} + \frac{i \lambda_r}{2} \, \theta'_{2r-1} \theta'_{2r} \right). \end{eqnarray} When we expand out the last expression in (\ref{fe}), each $\theta'_j$ occurs to power $0$ or $1$, and hence \begin{equation} \label{ra} \rho_A = \prod_{r=1}^m \left(\frac{1}{2} + \frac{i \lambda_r}{2} c'_{2r-1} c'_{2r} \right). \end{equation} From the product form (\ref{ra}) of $\rho_A$, we can immediately read off the entanglement spectrum.  Indeed, (\ref{ra}) shows that the density matrix $\rho_A$ decomposes into $m$ independent $2$ state systems.  The $r$'th one, where $r = 1, \ldots, m$, described by $1/2 + i \,(\lambda_r / 2)\, c'_{2r-1} c'_{2r}$, has eigenvalues $1/2 \pm \lambda_r/2$.  Hence the full entanglement spectrum is described by the set of $2^m$ Schmidt eigenvalues: \begin{equation}\label{sra} \left\{ \prod_{r=1}^m \left(\frac{1}{2} + s_r \frac{\lambda_r}{2} \right) \right\}_{s_r = \pm 1}. \end{equation}

\vspace{8pt}

\section{Consequences \label{cons}}
We have, in (\ref{sra}), computed the entire set of $2^m$ Schmidt eigenvalues comprising the entanglement spectrum of the ground state with respect to regions $A$ and $B$ in terms of the eigenvalues $\pm \lambda_r$.  These eigenvalues determine the edge mode spectrum of the corresponding spectrally flattened Hamiltonian (\ref{ha}).  Note that they only differ significantly from $\pm 1$ when the corresponding eigenstate is localized near the boundary of region $A$.  Also, if $\lambda_r \sim \pm 1$, which corresponds to a bulk mode, then $1/2 \pm \lambda_r / 2$ are close to $0$ and $1$; hence the contribution of such a mode to the entanglement becomes vanishingly small as $\lambda_r \rightarrow 1$ (or $\lambda_r \rightarrow -1$).  This just means that only edge modes contribute significantly to the entanglement.  Also, we see that a zero energy edge mode, corresponding to $\lambda_r=0$, is reflected as a nontrivial multiplicity of the entanglement spectrum (\ref{sra}).  Indeed, having $k$ such $\lambda_r=0$ results in a multiplicity of $2^k$ for the Schmidt eigenvalues.

We note the versatility of our Majorana fermion approach: our results hold not only for topological insulators, but for topological superconductors as well.  Furthermore, because the entanglement spectrum depends only on the ground state, our approach gives a way to diagnose topological order by looking at only the ground state wavefunction.  Indeed, we have proved that the entanglement spectrum can be reconstructed from the edge mode spectrum of the corresponding spectrally flattened Hamiltonian.  For a generic topological Hamiltonian with symmetry protected edge modes, we expect that the corresponding spectrally flattened Hamiltonian, being in the same phase, has the same symmetry protected edge modes.  Thus the entanglement spectrum is as good as the edge mode spectrum at discriminating between the topological phases.  However, the above statement can fail in special cases: for example, inversion symmetry can sometimes allow the spectrally flattened Hamiltonian to have a richer gapless edge mode structure than that possessed by the original Hamiltonian \cite{turner-2009}.  In fact, \cite{turner-2009} and \cite{pollmann-2009} interpret this discrepancy as a virtue, and argue that the entanglement spectrum is a more robust measure of topological order than the edge mode spectrum.

It would be very interesting to generalize this correspondence to interacting systems, as was done, for example, by Li and Haldane in \cite{haldane} for certain fractional quantum Hall states (see also \cite{thomale-2009}).

\acknowledgments I would like to acknowledge useful discussions with Alexei Kitaev, John Preskill, Gil Refael, Andrei Bernevig, and especially Jason Alicea.  This work was supported in part by the institute for Quantum Information under National Science Foundation grant no. PHY-0803371.

\bibliographystyle{aipproc-atitle}

\vspace{5mm} 
\bibliography{entspecrefs}

\begin{thebibliography}{21}
\expandafter\ifx\csname natexlab\endcsname\relax\def\natexlab#1{#1}\fi
\providecommand{\enquote}[1]{``#1''}
\expandafter\ifx\csname url\endcsname\relax
  \def\url#1{\texttt{#1}}\fi
\expandafter\ifx\csname urlprefix\endcsname\relax\def\urlprefix{URL }\fi
\providecommand{\eprint}[2][]{\url{#2}}

\bibitem[Kitaev and Preskill(2006)]{kp}
A.~Kitaev, and J.~Preskill, \enquote{Topological entanglement entropy,}
  \emph{Physical Review Letters} \textbf{96}, 110404 (2006).

\bibitem[Levin and Wen(2006)]{lw}
M.~Levin, and X.-G. Wen, \enquote{Detecting topological order in a ground state
  wave function,} \emph{Physical Review Letters} \textbf{96}, 110405 (2006).

\bibitem[Li and Haldane(2008)]{haldane}
H.~Li, and F.~D.~M. Haldane, \enquote{Entanglement Spectrum as a Generalization
  of Entanglement Entropy: Identification of Topological Order in Non-Abelian
  Fractional Quantum Hall Effect States,} \emph{Physical Review Letters}
  \textbf{101}, 010504 (2008).

\bibitem[Gioev and Klich(2006)]{klich}
D.~Gioev, and I.~Klich, \enquote{Entanglement entropy of fermions in any
  dimension and the Widom conjecture,} \emph{Physical Review Letters}
  \textbf{96}, 100503 (2006).

\bibitem[Schuch et~al.(2008)]{vers}
N.~Schuch, M.~M. Wolf, F.~Verstraete, and J.~I. Cirac, \enquote{Entropy scaling
  and simulability by Matrix Product States,} \emph{Physical Review Letters}
  \textbf{100}, 030504 (2008).

\bibitem[Hastings(2007)]{hastings}
M.~B. Hastings, An area law for one dimensional quantum systems (2007).

\bibitem[Kane and Mele(2005)]{KaneMele1}
C.~Kane, and E.~Mele, \enquote{$\mathbb{Z}_2$ Topological Order and the Quantum
  Spin Hall Effect,} \emph{Phys. Rev. Lett.} \textbf{95}, 146802 (2005).

\bibitem[Bernevig et~al.(2006)]{HgTe0}
B.~Bernevig, T.~Hughes, and S.-C. Zhang, \enquote{Quantum Spin Hall Effect and
  Topological Phase Transition in HgTe Quantum Wells,} \emph{Science}
  \textbf{314}, 1757 (2006).

\bibitem[Fu and Kane(2007)]{BiSb0}
L.~Fu, and C.~Kane, \enquote{Topological insulators with inversion symmetry,}
  \emph{Phys. Rev. B} \textbf{76}, 45302 (2007).

\bibitem[Hsieh et~al.(2008)]{BiSb1}
D.~Hsieh, D.~Qian, L.~Wray, Y.~Xia, Y.~Hor, R.~Cava, and M.~Hasan, \enquote{A
  topological Dirac insulator in a quantum spin Hall phase,} \emph{Nature}
  \textbf{452}, 970 (2008).

\bibitem[Klich and Levitov(2009)]{levitov}
I.~Klich, and L.~Levitov, \enquote{Many-Body Entanglement: a New Application of
  the Full Counting Statistics,} \emph{AIP CONFERENCE PROCEEDINGS}
  \textbf{1134}, 36 (2009).

\bibitem[Chung and Peschel(2001)]{peschel1}
M.-C. Chung, and I.~Peschel, \enquote{Density-Matrix Spectra of Solvable
  Fermionic Systems,} \emph{Physical Review B} \textbf{64}, 064412 (2001).

\bibitem[Peschel and Eisler(2009)]{peschel2}
I.~Peschel, and V.~Eisler, Reduced density matrices and entanglement entropy in
  free lattice models (2009), \urlprefix\url{arXiv.org:0906.1663}.

\bibitem[Barthel et~al.(2006)]{uli}
T.~Barthel, M.-C. Chung, and U.~Schollwoeck, \enquote{Entanglement scaling in
  critical two-dimensional fermionic and bosonic systems,} \emph{Physical
  Review A} \textbf{74}, 022329 (2006).

\bibitem[Kitaev(2006)]{kitaev}
A.~Kitaev, \enquote{Anyons in an exactly solved model and beyond,} \emph{Annals
  of Physics} \textbf{321}, 2 (2006).

\bibitem[Bravyi(2005)]{bravyi}
S.~Bravyi, \enquote{Lagrangian representation for fermionic linear optics,}
  \emph{QUANTUM INF.AND COMP.} \textbf{5}, 216 (2005).

\bibitem[Bray-Ali et~al.(2009)]{nba}
N.~Bray-Ali, L.~Ding, and S.~Haas, Topological order in paired states of
  fermions in two-dimensions with breaking of parity and time-reversal
  symmetries (2009), \urlprefix\url{arXiv:0905.2946}.

\bibitem[Turner et~al.(2009)]{turner-2009}
A.~M. Turner, Y.~Zhang, and A.~Vishwanath, Band topology of insulators via the
  entanglement spectrum (2009), \urlprefix\url{arXiv.org:0909.3119}.

\bibitem[Pollmann et~al.(2009)]{pollmann-2009}
F.~Pollmann, E.~Berg, A.~M. Turner, and M.~Oshikawa, Entanglement spectrum of a
  topological phase in one dimension (2009),
  \urlprefix\url{arXiv.org:0910.1811}.

\bibitem[Kitaev(2000)]{kitaev-2000}
A.~Kitaev, Unpaired majorana fermions in quantum wires (2000),
  \urlprefix\url{arXiv.org:cond-mat/0010440}.

\bibitem[Thomale et~al.(2009)]{thomale-2009}
R.~Thomale, A.~Sterdyniak, N.~Regnault, and B.~A. Bernevig, The entanglement
  gap and a new principle of adiabatic continuity (2009),
  \urlprefix\url{arXiv.org:0912.0523}.

\end{thebibliography}

\end{document}